\documentclass{elsart}
\usepackage{natbib,amssymb,latexsym,epsfig}
\begin{document}
\runauthor{Godbole and Pancheri}
\begin{flushright}
IISc-CTS/17/00\\
hep-ph/0101320
\end{flushright}

\begin{center}

{\large\bf    
$\gamma \gamma$ cross-sections and $\gamma \gamma$ colliders\footnote{Talk
presented by RMG at the International Workshop on High Energy Photon 
Colliders, DESY, Hamburg, June 2000.}} \\

\vskip 25pt

{\bf                        Rohini M. Godbole } \\ 

{\footnotesize\rm 
                      Centre for Theoretical Studies, 
                    Indian Institute of Science, Bangalore 560 012, India. \\ 
                     E-mail: rohini@cts.iisc.ernet.in  } \\ 

\bigskip

{\bf                       G. Pancheri } \\ 

{\footnotesize\rm 
                    Laboratori Nazionali di Frascati dell'INFN,  
                     Via E. Fermi 40, I 00044, Frascati, Italy. \\ 
                     E-mail: Giulia.Pancheri@lnf.infn.it  } \\

\vskip 30pt

{\bf                             Abstract 
}

\end{center}

\begin{quotation}
\noindent

We summarize the predictions of different models for total 
$\gamma \gamma$ cross-sections. The experimentaly observed rise 
of $\sigma_{\gamma \gamma}$ with $\sqrt{s_{\gamma \gamma}}$, faster 
than that for $\sigma_{\bar p p}$, $\sigma_{\gamma p}$ is in agreement 
with the predictions of the Eikonalized Minijet Models as opposed to 
those of the Regge-Pomeron models. We then show that a measurement 
of $\sigma_{\gamma \gamma }$ with an accuracy of 
$\lessapprox 8-9\% (6-7\%)$ is necessary to distinguish among 
different Regge-Pomeron type models (among the different 
parametrisations of the EMM models) and a  precision of 
$\lessapprox$ 20\% is required to distinguish among the predictions 
of the EMMs  and of those models which 
treat like 'photon like a proton',  for the energy 
range $300 < \sqrt{s_{\gamma \gamma}} < 500$ GeV. We
further show that the difference in model predictions for
$\sigma_{\gamma \gamma}$ of about a factor 2 at 
$\sqrt{s_{\gamma \gamma}}  = 700 $ GeV
reduces to $\sim$ 30\% when folded with bremsstrahlung 
$\gamma$ spectra to
calculate $\sigma ({e^+e^- \to e^+e^-\gamma \gamma  \to e^+e^-X})$. 
We point out then the special role that $\gamma \gamma$ colliders can 
play in shedding light on this
all-important issue of calculation of total hadronic cross-sections. 
\end{quotation}

\vskip 60pt
%%%%%%%%%%%%%%%%%%%%%%%%%%%%%%%%%%%%%%%%%%%%%%%%%%%%%%%%%%%%%%%%%%%%%
% ------------------Body of the Paper follows ----------------------%
%%%%%%%%%%%%%%%%%%%%%%%%%%%%%%%%%%%%%%%%%%%%%%%%%%%%%%%%%%%%%%%%%%%%%
% ------------------------------------------------------------------%
\newpage
\begin{frontmatter}
\title{$\gamma \gamma$ cross-sections and $\gamma \gamma$ colliders } 
\author[India]{Rohini M. Godbole} 
\author[Italy]{ G. Pancheri }
\address[India]{Centre for Theoretical Studies, Indian Institute of Science,
 Bangalore 560 012, India.}
\address[Italy]{Laboratori Nazionali di Frascati dell'INFN,
 Via E. Fermi 40, I 00044, Frascati, Italy.} 
\begin{abstract}
We summarize the predictions of different models for total $\gamma \gamma$
cross-sections. The experimentaly observed rise of $\sigma_{\gamma \gamma}$
with $\sqrt{s_{\gamma \gamma}}$, faster then that for $\sigma_{\bar p p}$,
$\sigma_{\gamma p}$ is in agreement with the predictions of the Eikonalized
Minijet Models as opposed to those of the Regge-Pomeron models. We then
show that a measurement of $\sigma_{\gamma \gamma }$ with an accuracy of
$\lessapprox 8-9\% (6-7\%)$ is necessary to distinguish among 
different Regge-Pomeron type models (among the different parametrisations of
the EMM models) and a  precision of $\lessapprox$ 20\% is required to 
distinguish among the predictions of the EMMs  and of those models which 
treat like 'photon like a proton',  for the energy 
range $300 < \sqrt{s_{\gamma \gamma}} < 500$ GeV. We
further show that the difference in model predictions for
$\sigma_{\gamma \gamma}$ of about a factor 2 at 
$\sqrt{s_{\gamma \gamma}}  = 700 $ GeV
reduces to $\sim$ 30\% when folded with bremsstrahlung $\gamma$ spectra to
calculate $\sigma ({e^+e^- \to e^+e^-\gamma \gamma  \to e^+e^-X})$. 
We point out then the special role that $\gamma \gamma$ colliders can 
play in shedding light on this
all-important issue of calculation of total hadronic cross-sections. 
%We end by giving an easy parameterization for calculation of 
%'low energy jets' at the $\gamma \gamma$ colliders.
\end{abstract}
\begin{keyword}
$\gamma \gamma$ cross-sections; Eikonalized Minijets;
 Regge-Pomeron
\end{keyword}
\end{frontmatter}

\section{ Introduction}

The subject of total $\gamma \gamma$ cross-section 
($\sigma^{\rm tot}_{\gamma \gamma}$) is a very
important one, both from a theoretical point of view of understanding
calculation of total/inelastic hadronic cross-sections and a much more
pragmatic one of being able to predict the hadronic backgrounds~\cite{0}
at the future linear colliders~\cite{1} due to $\gamma \gamma$ processes. 
The recent data on energy dependence of $\sigma_{\gamma p}$ and 
$\sigma_{\gamma \gamma}$ available from HERA~\cite{2,2'} and LEP~\cite{3,4}
respectively,  have established that these cross-sections
rise with energy. They have provided us with an additional laboratory to
test/develope the models for calculation of total hadronic
cross-sections~\cite{5}. However 
$\sigma^{tot}_{\gamma \gamma}$  and $\sigma^{tot}_{\gamma p}$ 
is measured by studying the reactions
$e^+e^- \to e^+e^-\gamma \gamma  \to e^+e^-X $ and $ep \to e\gamma p \to eX$
respectively.  The unfolding of $\gamma \gamma$ cross-sections from the 
measured ${e^+e^-}$ cross-sections is a major source of error in 
the measurement of $\sigma^{\rm tot}_{\gamma \gamma}$ . This is 
exemplified by the dependence of $\sigma^{tot}_{\gamma \gamma}$ 
presented by LEP collaborations on the Monte
Carlo used for unfolding; the difference in the normalization of the
extracted cross-sections using different Monte Carlos can be as much as 
50\% at the high energy end~\cite{4} and can be seen in the data shown in
Fig.~\ref{figtwo} later. Hence a $\gamma \gamma$ collider with
$\sqrt{s_{\gamma \gamma}}$ in the range 300-500 GeV will provide an opportunity
for an unambiguous and accurate measurement of 
$\sigma^{\rm tot}_{\gamma \gamma}$.
With such data, we will have information for 
$\gamma p$, $\gamma \gamma$ and ${pp(p \bar p)}$ 
for similar range of $\sqrt{s}$ values. Such information will
undoubtedly provide important pointers to arrive at a better theoretical
understanding, from first principles, of total/inelastic cross-sections of
hadronic processes. 
\begin{figure}
\begin{center}
\includegraphics*[scale=0.5]{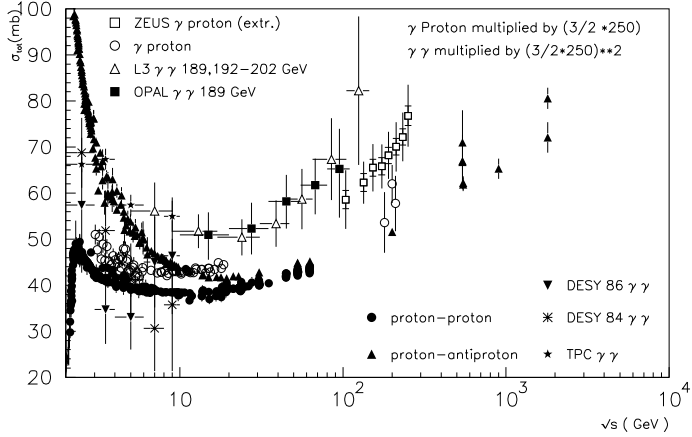}
\caption{Energy dependence of $\sigma^{\rm tot}_{ab}$ for various 
choices of $a,b$ as indicated in the figure.}
\label{fone}
\end{center}
\end{figure}
Fig.~\ref{fone} shows the $\sigma^{\rm tot}_{pp/\bar p p }$,
$(3/2) 250~\sigma^{\rm tot}_{\gamma p}$ and $((3/2) 250 )^2$
$\sigma^{\rm tot}_{\gamma \gamma}$ in the same graph. The multiplication 
factors are guided by simple VMD considerations. In this figure we have 
included the latest L3 data from LEP-II~\cite{15}.
We see in the figure that the available
data show indications of somewhat higher rate of rise with energy for
$\sigma^{\rm tot}_{\gamma \gamma} (\sigma^{\rm tot}_{\gamma p})$ 
as compared to $\sigma^{tot}_{pp/p\bar p}$. Hence 
$\sigma^{tot}_{\gamma \gamma}$ will be an
important quantity to be measured accurately at the future $\gamma \gamma$
colliders. In this note, we assess the success of various models for
$\gamma \gamma$ cross-sections, in `explaining' currently available data 
and point out the precision necessary to be able to distinguish between 
different models~\cite{6}.

\section{Theoretical Models : }

 There are two different classes of models used to calculate the $\gamma
\gamma$ cross-sections.

1] Models which treat a photon like a proton: these models obtain the 
$\gamma \gamma$ total cross sections through extrapolations of some or all
of the proton properties. There exist three different types.
\begin{description}
\item{(a)} Regge/Pomeron type models where the (increase) decrease of the
cross-sections with energy is given by the (Pomeron) Regge part. 
These models assume
factorization of residues at the pole. The total cross-section is written as
\begin{equation}
\sigma_{ab}^{\rm tot} = Y_{ab} {s}^{- \eta} + X_{ab} {s}^{\epsilon}.
\end{equation}
The coeffecients X,Y for the $\gamma \gamma$ case are 
determined~\cite{7} by using the fitted values of X,Y for the 
$pp({\bar p p)}$ and $\gamma p$ case. 
A somewhat more complicated model~\cite{8} gives similar predictions.

\item{(b)} In a model by C. Bourelly et al, \cite{9} $\sigma_{\gamma \gamma}$ 
is obtained by a straightforward scaling of $\sigma_{pp}$ {\it viz.}, 
$\sigma_{\gamma \gamma}$ = A$\sigma_{pp}$

\item{(c)}A model by Badelek and colalborators~\cite{10}(BKKS) again 
presents an extrapolation of the knowledge on $\sigma_{pp}$ coupled with 
VMD ideas. They fit the parameters by using data on $\sigma_{pp}$ and 
then make predictions for $\sigma_{\gamma \gamma}$.  
\end{description}

2] The second type of models are the QCD based/inspired models. In this
case, the rise of the cross-sections with energy is driven by the rise in
production of small transverse momentum jets in hadronic collisions. In the
case of (say)$\gamma p$ collisions, $\sigma^{\rm tot}_{\gamma p}$ is given by
\begin{equation} 
\sigma^{\rm tot}_{\gamma p} = 2 P^{\rm had}_{\gamma p} 
\int d^2\vec{b} [1 - e^{-\chi^{\gamma p}_I} cos \chi^{\gamma p}_R]
\label{two}
\end{equation}
where $P^{\rm had}_{\gamma p}$ is the hadronization probability for a photon
given by
\begin{equation}
P^{\rm had}_{\gamma p} = P^{\rm had} = \sum_{V=\rho,\omega,\phi} 
\frac{4 \pi \alpha}{f_V^2} \simeq \frac{1}{240}.
\end{equation}
and $\chi^{\gamma p}_R$ = 0. 
Different models using the minijet idea differ in their
choices of the imaginary part of the eikonal $\chi_I$.
While calculating the total/elastic/inelastic cross-sections for the
case of $pp/\bar p p$, $P_{\gamma p}^{\rm had}$ in Eq.~\ref{two}, is replaced 
by unity and for the case of $\gamma \gamma$ collisions by $(P^{\rm had})^2$
respectively.
\begin{description}
\item{(a)} 
For the eikonalized minijet model EMM~\cite{11} we have
\begin{equation} 
2 \chi^{\gamma p}_I(s,b) = A_{\gamma p}(b) \left[\sigma^{\rm soft}_{\gamma p}(s)
+ {\frac{1}{P^{\rm had}}}\sigma^{\rm jet}_{\gamma p}(s,p_{Tmin})\right]
\end{equation}
Here $A_{\gamma p}(b)$ is the overlap function in the transverse space 
for the partons in colliding hadrons, $\sigma^{\rm soft}_{\gamma p}$ 
is the nonperturbative parameter describing the soft contribution to 
the cross-section and it is of the order of typical hadronic cross-sections. 
$\sigma^{\rm jet}_{\gamma p}$ is the hard jet cross-section obtained by 
integrating the usual jet cross-sections for $\gamma p$ collisions from 
a lower cut-off on $p_T$: $p_{Tmin}$. $A_{\gamma p}(b)$ here is modelled 
in terms of the Fourier Transform of the form factors or that of the 
measured transverse momentum distribution of partons in the photon and proton.

Once the various parameters are fitted using $\gamma p$ data, the 
corresponding parameters for the $\gamma \gamma$ case are obtained assuming
$$
\sigma_{\gamma \gamma}^{\rm soft} = \frac{2}{3} \sigma_{\gamma p}^{\rm soft}, 
P_{\gamma \gamma}^{\rm had} = (P^{\rm had})^2.
$$ 
All the rest of the quantities are defined similar to  the $\gamma p$ case.

\item{(b)} 
In another formulation of the EMM~\cite{12}, one calculates  A(b)
in terms of transverse momentum distribution for the partons.
However, instead of using the experimentally measured trnsaverse 
momentum distribution,  one calculates it in terms of soft gluon emission 
from the initial state valence quarks. This has the advantage of being 
able to produce also the initial fall of the 
$\sigma_{\gamma \gamma}^{\rm tot}$ with energy at low energies.

\item{(c)} In a third QCD based model~\cite{13} the eikonal as 
well as the overlap function are obtained by using factorization and 
simple scaling from the $pp$ case. The imaginary part of the eikonal in 
this case is given by 
\begin{equation}
\chi_{I} = P_{gg} + P_{gq} + P_{qg}.
\end{equation}
with 
$P_{ij}  =  W_{ij} (b,\mu_{ij}) \sigma_{ij} (s)$ where 
$W(b,\mu)$ for each case given by,
\begin{equation}
W (b,\mu) = \int \frac{d^2\vec q}{(2\pi)^2} e^{i\vec b \cdot \vec q} 
|F(q)|^2.
\end{equation}
$F(q)$ here is taken to be the dipole form factor. The various
parameters $\mu_{ij}$ and $\sigma_{ij}$ are fitted to the $pp/ p \bar p$
data.  The corresponding ones for the $\gamma p$ and $\gamma \gamma$ case 
are then determined by simple scaling arguments implied by the Quark-Parton
model. 
\end{description}

\section{Predictions of various models for $\sigma_{\gamma \gamma}$ : }

\begin{figure}
\begin{center}
\includegraphics*[scale=0.4]{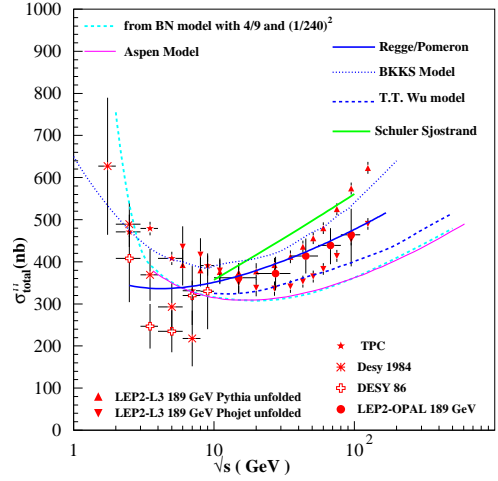}
\caption{Predictions of models which treat photon like a proton}
\label{figtwo}
\end{center}
\end{figure}

\begin{figure}
\begin{center}
\includegraphics*[scale=0.4]{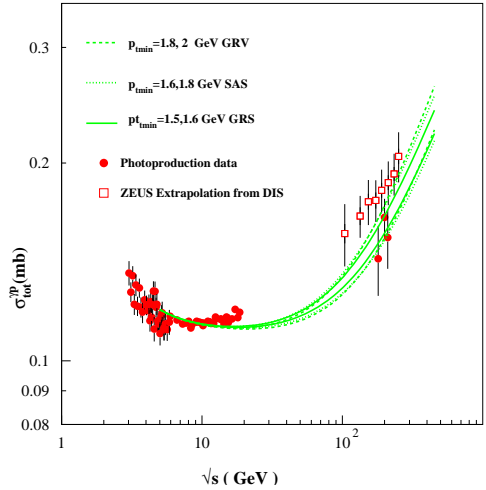}
\caption{ EMM predictions for $\sigma_{\gamma p}$ with different densities 
for the photonic partons~\protect\cite{8,GRV,GRS}.}
\label{figthree}
\end{center}
\end{figure}

\begin{figure}
\begin{center}
\includegraphics*[scale=0.4]{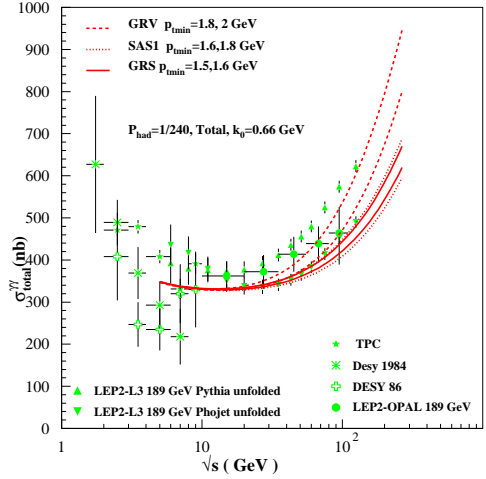}
\caption{ Comparison of the older LEP2 data with predictions of the EMM 
model in the total formulation for different photonic parton densities.}
\label{figfour}
\end{center}
\end{figure}

\begin{figure}
\begin{center}
\includegraphics*[scale=0.5]{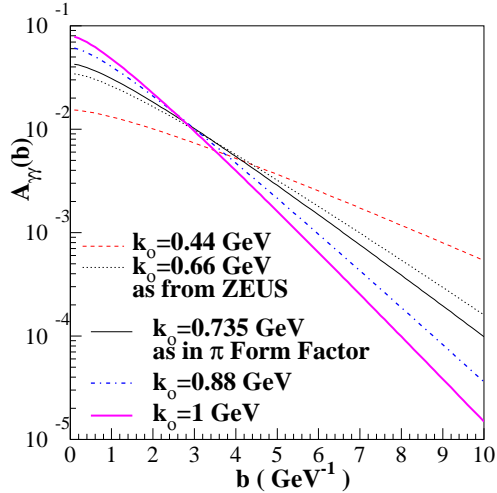}
\caption{$A_{\gamma \gamma}$ for different values of $k_0$. }
\label{figfive}
\end{center}
\end{figure}

\begin{figure}[hbt]
\begin{center}
\includegraphics*[scale=0.4]{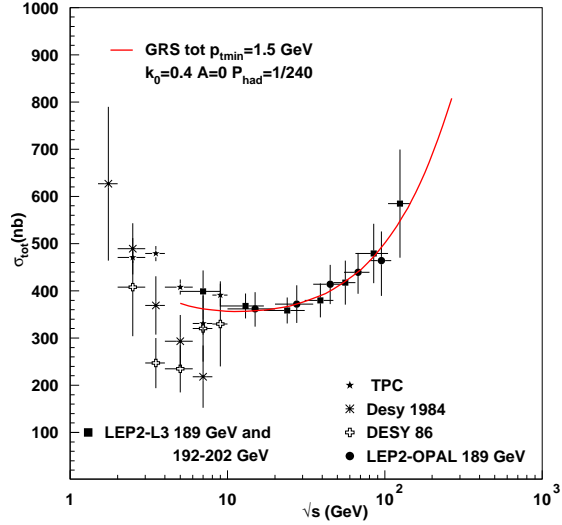}
\caption{Latest LEP2 data along with EMM prediction in total formulation
with $k_0$ = 0.4}
\label{figsix}
\end{center}
\end{figure}

\begin{figure}
\begin{center}
\includegraphics*[scale=0.4]{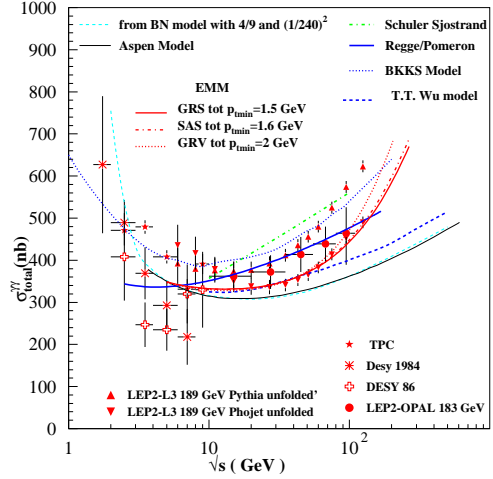}
\caption{A comparison of all the various model predictions with the latest 
data.}
\label{figseven}
\end{center}
\end{figure}

Fig.~\ref{figtwo} shows a comparison of the current data with the prediction of
various `photon-like a proton' models. As one can see, all these models have
some difficulty producing the faster rise shown by the data for
$\sigma_{\gamma \gamma}$.
Here we have included the predictions of two QCD based
models, the BN model~\cite{12} and  the Aspen Model~\cite{13}, as well. 
We do see that the BN model does quite well with the fall at low energies
as well. In Fig.~\ref{figthree} we compare the predictions of the 
EMM model in the total formulation for $\sigma_{\gamma p}$, with the data.
Note that the newer data
on  $\sigma^{tot}_{\gamma p}$ obtained by the extrapolation of the DIS data
to photoproduction limit~\cite{2'} lies consistently above the
$\sigma^{tot}_{\gamma p}$ measured previously~\cite{2}. The A(b) here is
modelled as the Fourier Transform of the product of electromagnetic form 
factors for the
proton and the experimentally measured transverse momentum distribution for
photonic partons. Here we have used the central value of the parameter $k_0$ =
0.66 where the transverse momentum distribution is measured~\cite{14} 
to be $\propto$ $1/(k_T^2 + k_0^2)$.
Having fixed all the values by $\sigma^{\rm tot}_{\gamma p}$, if we now
calculate $\sigma^{\rm tot}_{\gamma \gamma}$ we get predictions shown in 
Fig.~\ref{figfour}. 
In this figure we have the $189$ GeV data from the OPAL~\cite{3} and
L3~\cite{4} collaborations.
Fig.~\ref{figfive} shows the $A_{\gamma \gamma}(b)$ \cite{11} 
for different values of $k_0$, allowed by the experimental 
measurement~\cite{14} of $k_0 = 0.66 \pm 0.22$.
As $k_0$ decreases (increases), the curves in Fig.~\ref{figfour} 
will move up (down). Actually Fig.~\ref{figsix} shows the 
prediction of the EMM model using $k_0 = 0.4$ along
with the same OPAL data as in Fig.~\ref{figfour} and the latest 
L3 data~\cite{15}. Values of all the other parameters which have been
used are as given in the figure. We see that the EMM model is able to produce 
the trend of the faster rise quite well.  Fig. ~\ref{figseven} shows a 
comparison of all the model predictions with each other and
the data. We notice that the rate of rise of total cross-sections in the
EMM/BKKS models is quite different from those in Regge-Pomeron type models.

%\begin{minipage}[t][4in]
\begin{table}
\begin{center}
\caption{Predictions for different `proton-like' models}
\begin{tabular}{|ccccc|}
\hline 
$\sqrt{s_{\gamma \gamma}} (GeV)$ & Aspen &  T.T. Wu & DL & $1 \sigma$ \\ \hline
\hline
 20    & 309 nb & 330 nb & 379 nb &  7\%  \\ \hline
 50    & 330 nb & 368 nb & 430 nb &  11\%   \\ \hline
 100   & 362 nb & 401 nb & 477 nb &  10\%   \\  \hline
 200   & 404 nb & 441 nb & 531 nb &  9\%   \\  \hline
 500   & 474 nb & 515 nb & 612 nb &  8\%   \\  \hline
 700   & 503 nb & 543 nb & 645 nb &  8\%   \\ \hline
\end{tabular}
\label{table1}
\end{center}
\end{table}
%\end{minipage}
%\~~~\
%\begin{minipage}[t][4in]
\begin{table}
\begin{center}
\caption{Predictions for different QCD based models.}
\begin{tabular}{|ccccc|}
\hline 
$\sqrt{s_{\gamma \gamma}} $ &EMM,Inel,GRS &EMM,Tot,GRV & BKKS& $1 \sigma$ \\ 
& ($p_{tmin}$=1.5 GeV)& ($p_{tmin}$=2 GeV)              & GRV & \\ \hline
\hline
 20    &399  nb & 331 nb      & 408 nb &   2 \%  \\ \hline
 50    &429  nb & 374 nb      & 471 nb &   9\%   \\ \hline
 100   &486  nb & 472 nb      & 543 nb &   11\%   \\  \hline
 200   & 596 nb & 676 nb      & 635 nb&   6\%   \\  \hline
 500   & 850 nb & 1165 nb      & 792 nb &  7  \%   \\  \hline
 700   & 978 nb & 1407 nb     & 860 nb &   13 \%   \\ \hline
\end{tabular}
\label{table2}
\end{center}
\end{table}
%\end{minipage}
The tables 1 and 2, give the precision with which 
$\sigma_{\gamma \gamma}$ needs to be measured, at the $\gamma \gamma$ 
colliders with $\sqrt{s_{\gamma \gamma}}$ in the range of 300-500 GeV,
to be able to distinguish between the different 'photon is like a proton' 
models as well as the EMM/BKKS models. As we can see, a precision
of $\lessapprox 7\% $ is required to distinguish among the different 
'photon like a proton' models from one another, whereas only a precision of 
$\lessapprox 20\%$ 
is required to distinguish these predictions from those of the QCD
based/inspired models which tend to predict a faster rise, in the energy range
$300 < \sqrt{s_{\gamma \gamma}} < 500$ GeV. With $\gamma \gamma$ cm energy
$\approx 700 $ GeV, the difference between the predictions of the
Aspen~\cite{13} and EMM total formulation~\cite{6} can be as large
as a factor of 2.

\begin{figure}
\begin{center}
\includegraphics*[scale=0.4]{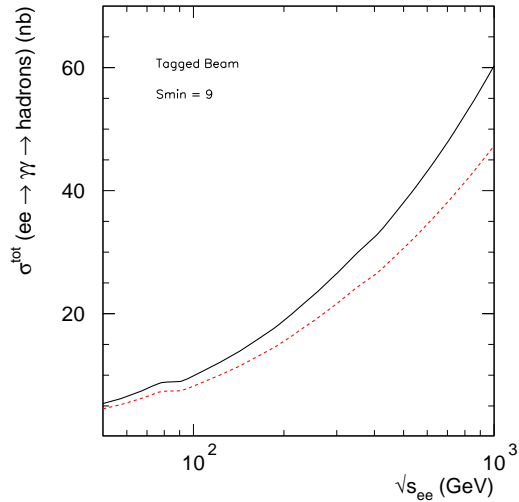}
\caption{Total hadron production cross-section via two-photon
processes at $e^+e^-$ colliders.}
\label{figeight}
\end{center}
\end{figure}

However, when these $\gamma \gamma$ cross-sections are convoluted with the
spectrum of the bremsstrahlung photons to calculate $\sigma$ ($e^+e^- \to
e^+e^-{\gamma \gamma} \to e^+e^-X)$ using the WW approximation, we find 
that these big differences get reduced to about  $30 \%$. this is shown in
Fig.~\ref{figeight}. This demonstrates the much more superior role that 
the $\gamma \gamma$ colliders can play in deciding which is the right 
theoretical framework for calculation of total cross-sections .
\section{Conclusions}
Thus in conclusion we can say the following
\begin{enumerate}
\item 'Photon is like a proton' models predict a rise of $\sigma_{\gamma
\gamma}$ with $\sqrt{s_{\gamma \gamma}}$, slower than shown by the data; 
i.e. value of predicted $\epsilon$ is lower than what the data seem to show.
\item The extrapolated $\gamma p$ data seem to show similar trends.
\item The predictions of the EMM model show good agreement with the data.
\item Even in the EMM formulations use of Bloch Nordsieck ideas to calculate
the overlap function $A(b)$ seems to slow down this rise.
\item An obvious improvement in the EMM models is to try and determine
$A(b)$ by more refined `theoretical' ideas or determine it in terms of the
multiple parton interactions measured  at the HERA/Tevatron collider.
\item However, extraction of $\sigma_{\gamma \gamma}$ and $\sigma_{\gamma p}$ 
from $\sigma_{e^+e^-}$ and $\sigma_{ep}$ respectively, is no mean task and 
has large uncertainties.
Moreover, a difference of about a factor two in the predicted values of 
$\sigma^{\rm tot}_{\gamma \gamma}$ in different models, gets reduced to only
about $30\%$ when folded with the photon spectrum expected in the 
WW approximation in $e^+e^-$ collisions. While the good part is that it reduces
the uncertainty in our predictions of the hadronic background at the $e^+e^-$
linear colliders due to the corresponding uncertainties in 
$\sigma^{\rm tot}_{\gamma \gamma}$, the studies of  two-photon hadronic 
cross-sections at $e^+e^-$ colliders, will not be very efficient in 
shedding  much light on the theortical models used to calculate them.
\item Therefore measurements of total cross-sections at a 
$\gamma \gamma$ collider with its monochromatic photon beam, in the 
energy range $300 < \sqrt{s_{\gamma \gamma}} < 500$ GeV, 
can play a very useful role 
in furthering our understanding of the 'high' energy photon interactions.
A precision of $\lessapprox 7-8\%(8-9\%)$ 
is required to distinguish among the different formulations of the EMM models 
(models which treat photon like a proton), where as a precision of 
$\lessapprox 20\%$ is required to distinguish betwen these two types of models.
\end{enumerate}

\section{Acknowledgements:}
We grateful to A. de Roeck for the suggestion to study the issue of
precision of measurement at the linear colliders.

\end{document}